# The Clock Distribution System for the ATLAS Liquid Argon Calorimeter Phase-I Upgrade Demonstrator

Binwei Deng,[a,b] Hucheng Chen,[c] Kai Chen,[c] Jinghong Chen,[d,e] Datao Gong,[b] Di Guo,[b,f] Xueye Hu,[c] Deping Huang,[e] James Kierstead,[c] Xiaoting Li,[b,g] Chonghan Liu,[b] Tiankuan Liu,[b,*] Annie C. Xiang,[b] Hao Xu,[c] Tongye Xu,[h] Yang You,[e] Jingbo Ye[b]

[a] *School of Electric and Electronic Information Engineering, Hubei Polytechnic University,*
   *Huangshi, Hubei 435003, P. R. China*

[b] *Department of Physics, Southern Methodist University,*
   *Dallas, TX 75275, USA*

[c] *Department of Physics, Brookhaven National Laboratory,*
   *Upton, NY 11973, USA*

[d] *Department of Electrical Engineering, University of Houston*
   *Houston, Texas, 77004*

[e] *Department of Electrical Engineering, Southern Methodist University,*
   *Dallas, TX 75275, USA*

[f] *Department of Modern Physics, University of Science and Technology of China,*
   *Hefei, Anhui 230026, P. R. China*

[g] *Department of Physics, Central China Normal University,*
   *Wuhan, Hubei 430079, P.R. China*

[h] *MOE Key Lab on Particle Physics and Particle Irradiation, Shandong University,*
   *Ji'nan, Shandong 250100,P.R. China*
   E-mail: tliu@mail.smu.edu

ABSTRACT: A prototype Liquid-argon Trigger Digitizer Board (LTDB), called the LTDB Demonstrator, has been developed to demonstrate the functions of the ATLAS Liquid Argon Calorimeter Phase-I trigger electronics upgrade. Forty Analog-to-Digital converters and four FPGAs with embedded multi-gigabit-transceivers on each Demonstrator need high quality clocks. A clock distribution system based on commercial components has been developed for the Demonstrator. The design of the clock distribution system is presented. The performance of the clock distribution system has been evaluated. The components used in the clock distribution system have been qualified to meet radiation tolerance requirements of the Demonstrator.

# Contents

## 1. Introduction

The ATLAS Liquid Argon Calorimeter (LAr) trigger electronics will be upgraded in Phase-I to improve the trigger performance of the ATLAS experiment [1]. A LAr Trigger Digitizer Board (LTDB), which samples and digitizes up to 320 detector channels at 40 MHz and 12-bit resolution and transmits all the data off the detector through 40 optical fibers, will be developed. The LTDB will operate in a harsh radiation environment [2]. The LTDB will be implemented using radiation-tolerant components such as GBTX [3]. In order to demonstrate the full function of the LTDB, a prototype board called the Demonstrator, which is based on Commercial-Off-The-Shelf (COTS) components, has been developed. Each Demonstrator includes 40 octal-channel Analog-to-Digital converters (ADCs) (part number ADS5272 produced by Texas Instruments [4]) and 4 FPGAs (Xilinx Kintex 7 series, part number XC7K325TFFG900). Each ADC needs a 40-MHz LVTTL clock. Each FPGA, which is used to transmit detector data out of the detector, has 16 embedded multi-gigabit-transceivers (GTXs) and needs 2 160-MHz clocks. Two out of four FPGAs need two extra clocks for system monitoring. Therefore, it is critical to distribute high-quality clocks to all ADCs and FPGAs. In this paper, we present a clock distribution system for the LTDB Demonstrator. The design, verification, and radiation qualification of the clock distribution system are discussed.

## 2. Design and Verification of the clock distribution system

The block diagram of the clock distribution system for the Demonstrator is shown in figure 1. The clock source of the Demonstrator comes from a Gigabit Link Interface Board (GLIB) [5] located on the back-end counting room via two optical fibers. A quad-channel QSFP optical receiver (part number AFBR-79EIDZ produced by Avago Technologies) converts the optical signals to the electronic signals and sends the electrical signals to two Kintex-7 FPGAs. Each of the two FPGAs has one GTX bank which runs the GBT firmware [6] and recovers two 120-MHz clocks from the 4.8-Gbps data. Each of the two recovered 120-MHz clocks is sent into a jitter cleaner, LMK03200 or CDCE62005 (all produced by Texas Instruments). Both LMK03200 and CDCE62005 are also frequency synthesizers which generate a 40-MHz LVDS clock and a 160-MHz LVDS clock. The 40-MHz clocks generated by either LMK03200 or CDCE62005 are fanned out to 40 ADCs by two stages of ten-output low-jitter clock buffers CDCLVC1310. The last stages of clock buffers CDCLVC1310 also convert the LVDS signals to LVTTL clocks. The 160-MHz clocks generated by either LMK03200 or CDCE62005 are fanned out to the 4 FPGAs by a 12-output low-jitter LVDS clock buffer CDCLVD1212. Each of the 4 FPGAs uses 10 GTXs to transmit the detector data through a 12-channel (only 10 channels are used) pluggable, parallel-fiber optics module and 10 optical fibers (not shown in the figure). The left FPGA and the right FPGA use two of the four output fibers of the QSFP to monitor the Demonstrator status (not shown in the figure). Each of these two FPGAs has one extra 160-MHz clock from CDCLVD1212.

In order to improve the reliability, two FPGAs and two jitter cleaners are redundantly used to recover the 160-MHz clocks from two optical fibers. Either CDCLVC1310 or CDCLVD1212 has two inputs which can be remotely configured to select one clock source.

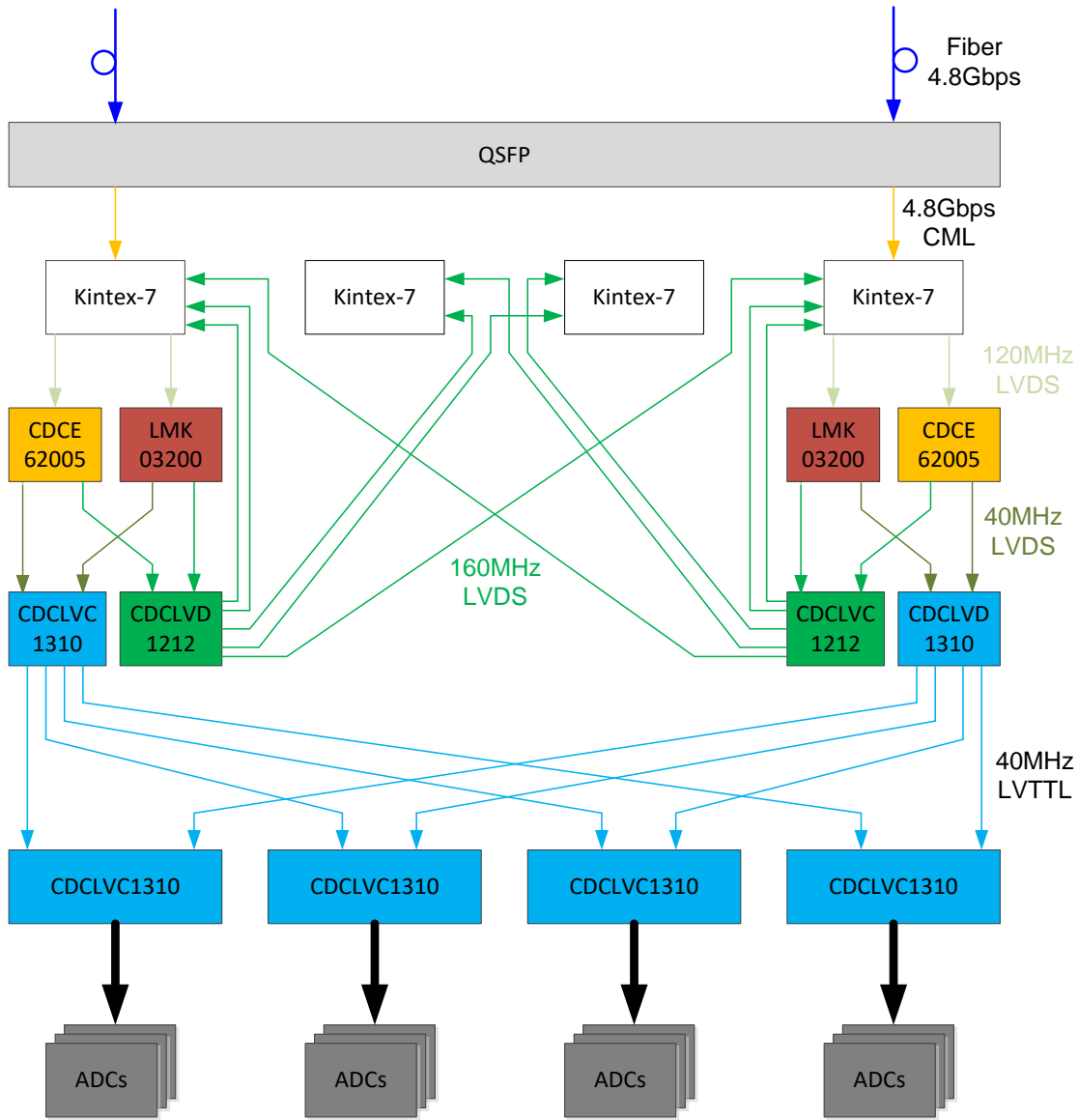

Figure 1: The block diagram of the clock distribution system

The performance of the clock distribution system has been evaluated. All components, including the optical transceiver, the FPGA, the jitter cleaners, and the clock buffers, have been tested independently. Both CDCE62005 and LMK03200 have been tested to generate 40-MHz and 160-MHz clocks from a 120-MHz clock. Table 1 lists the jitter performance of an FPGA recovered clock before and after a jitter cleaner is applied. It can be seen that both random jitter and periodic jitter, after the jitter cleaner, are less than those before the jitter cleaner.

Table 1: The jitter measurement summary

| Clock jitter | | Random (RMS, ps) | periodic (pk-pk, ps) |
|---|---|---|---|
| Before cleaner | | 5.9 | 81.2 |
| After cleaner | CDCE62005 | 5.1 | 6.5 |
| | LMK03200 | 1.4 | 12.9 |

The clock distribution system was verified in a Kintex-7 Evaluation Kit (Xilinx Part number KC705) running real data transmission firmware. The block diagram of the verification system is shown in figure 2. The 16 GTXs of the Kintex-7 FPGA are grouped in 4 banks (Banks 115-118). An Evaluation Board of the clock synthesizer Si5338 generated a 120-MHz clock and was used as the reference clock of the GTX Bank 118. The GTX Bank 118 ran the GBT FPGA firmware [6] at 4.8 Gbps and the transmitter was looped back to the receiver to recover a 120-MHz clock. The recovered clock was sent to another clock synthesizer Si5338. The clock synthesizer generated a 320-MHz (the clock frequency was changed later on the LTDB Demonstrator to 160 MHz) clock which was used as the reference clock of the GTX Banks 116 and 117. Each of the 8 GTX channels in Banks 116 and 117 ran the 5.12-Gbps firmware which emulated the data transmission scheme [7] being developed for the ATLAS LAr Phase-I upgrade.

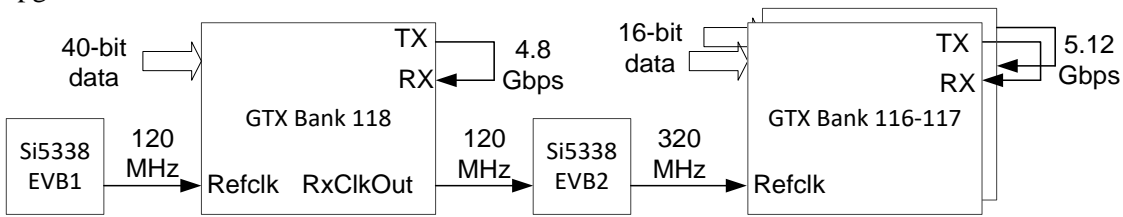

Figure 2: The block diagram of the clock distribution system

The clock distribution system has been used in the LTDB Demonstrator which has been installed in a spare slot of a Front End Crate (I06) of the EM Barrel Calorimeter.

## 3. Radiation Tolerance Qualification

After the Demonstrator is installed, it will operate at a low luminosity (100 fb$^{-1}$ in 3 years) for one to three years (2015-2017). Therefore, the radiation level for the Demonstrator is lower than that for the LDTB. However, since the Demonstrator is based on COTS components, the radiation tolerance of all components has to be evaluated. The simulated radiation level [2, 4] at the position where the LTDB Demonstrator is installed, safety factors [8], and radiation tolerance criteria [2] are listed in table 2. Total Ionizing Dose (TID), Non-Ionizing Energy Loss (NIEL), and Single-Event Effects (SEEs) are included.

The components, including optical transceivers, clock buffers, jitter cleaners & clock synthesizers, and the FPGA, were irradiated for TID and SEEs. We have not tested any

– 3 –

component for Non-Ionizing Energy Loss (NIEL) (the QSFP was tested in a proton beam, but the fluence was not high enough to qualify the module). Most modern integrated circuits, including the FPGA, clock buffers, and jitter cleaners, are pure CMOS circuits which are naturally insensitive to NIEL [2]. Previous tests [9] have shown that VCSELs and photodiodes degrade little at the low fluence of $1.6\times10^{12}$ cm$^{-2}$. Several types of optical fibers have been qualified before [10-11]. The irradiation test setup of the QSFP optical transceiver is discussed in [12]. Since most of the components are related to clocks, we take the clock buffers as the example to explain how we tested.

Table 2: the simulated radiation level, safety factors, and radiation tolerance criteria

|  | Simulated radiation level | Safety factor | | | Radiation tolerance criteria (3 years) |
|---|---|---|---|---|---|
|  |  | Simulation | Low dose rate effect | Lot-lot variation |  |
| TID | 3.0 Gy(SiO2) | 1.5 | 5 | 4 | 90 Gy(SiO2) |
| NIEL | $2.0\times10^{11}$ cm$^{-2}$ 1-MeV equ. neutrons in Si | 2 | 1 | 4 | $1.6\times10^{12}$ cm$^{-2}$ neutrons |
| SEE | $2.8\times10^{10}$ cm$^{-2}$ >20 MeV hadrons | 2 | 1 | 4 | $2.3\times10^{11}$ cm$^{-2}$ hadrons |

The TID test of LMK03200 was performed in x-rays to the total dose of 145 Gy(SiO2) with an in-house x-ray biological irradiator (part number X-RAD iR-160 produced by Precision X-ray, Inc.). The maximum x-ray energy is 160 keV. The peak x-ray energy is about 30 keV with a 2-mm aluminum filter. During the test, the power consumption and the output amplitude was monitored. The test lasted 5 hours. During the test, the power consumption changed less than 1%. After the irradiation, the output amplitude increased 1%.

Figure 3 is the SEE test setup of the clock jitter cleaner LMK03200. A clock generator evaluation board (part number Si5338-EVB produced by Silicon Laboratories Inc.) generated a 100-MHz clock and a 50-MHz clock. The 50 MHz clock went through the clock jitter cleaner. An Altera Stratix II GX FPGA used the 100-MHz clock to sample the cleaned 50-MHz clock. At the normal operation state, for each 50 MHz, the FPGA should get an alternative sequence (such as "010101…"). If there was disturbance from such an alternative sequence, an error would be recorded. During the test the jitter cleaner was put inside a neutron beam, whereas the other components were set aside away from the beam. The SEE test of LMK03200 was done at Los Alamos Neutron Science Center (LANSCE) of Los Alamos National Laboratory. Figure 4 is a picture of the test setup at LANSCE. The maximum energy was 800 MeV. The energy spectrum of neutrons is similar to that simulated in the ATLAS application environment [4]. The test lasted about 1.8 days (valid beam time about 1.6 days). The flux was in average $2.94\times10^5$ cm$^{-2}$s$^{-1}$. The total fluence was $3.97\times10^{10}$ cm$^{-2}$. During the test, 24 errors were observed. Each of the errors lasted for hundreds of milliseconds and recovered automatically afterwards. Those errors were treated as Single-Event Functional Interrupts (SEFIs). The SEFI cross section was estimated to be $6.1\times10^{-10}$ cm$^2$/device. No single Single-Event Upset (SEU) was observed. The SEU cross section was estimated to be less than $2.5\times10^{-11}$ cm$^2$/device.

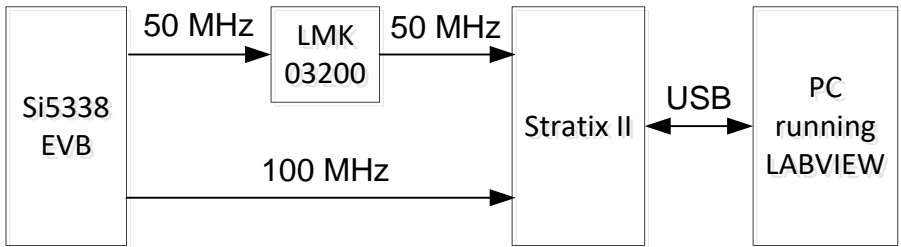

Figure 3: the SEE test setup of clock buffers and jitter cleaner

The TID test results are summarized in Table 3. All components can operate in the ATLAS environment for at least 2.7 years (the QSFP optical transceiver) or longer than three years (all the others).

The SEE test results are summarized in Table 4. The error rate of each component in the application operation environment is estimated. The upper limits of cross section and the estimated error rate are listed in the table if no error was observed during the test. All components have less than one error per week. The test results show that all components meet the radiation tolerance requirements.

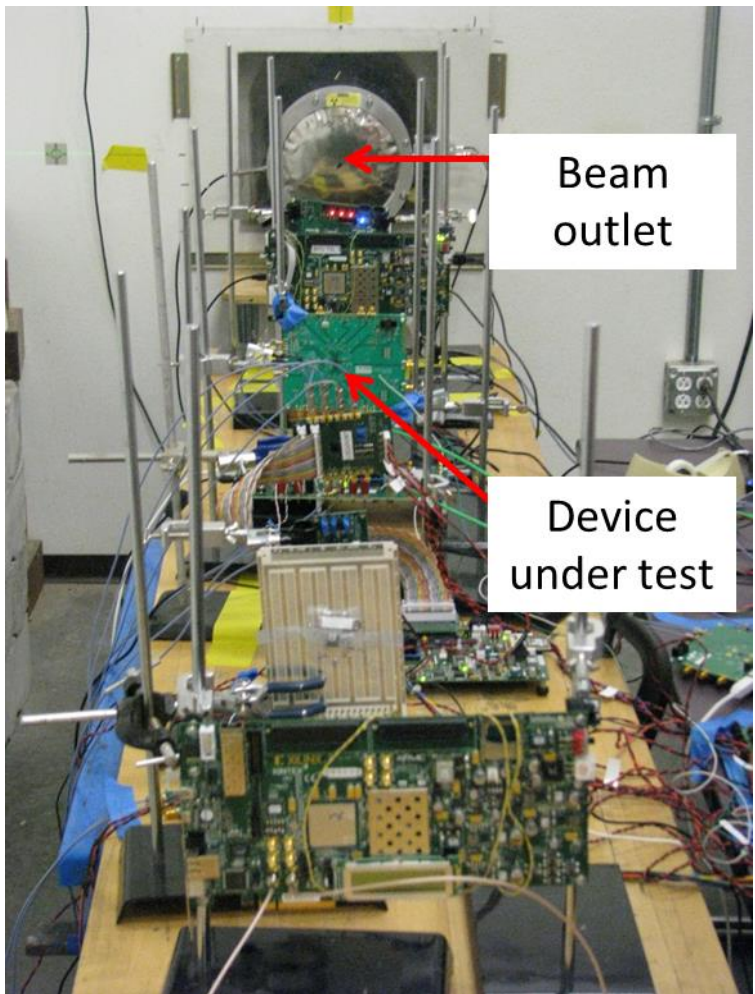

Figure 4: picture of the SEE test setup of clock buffers

Table 3 The TID test result summary

| Device | Part number | Radiation | TID (Gy) | Degradation | Max power current change |
|---|---|---|---|---|---|
| QSFP optical transceiver | AFBR-79EIDZ | X-ray | 81 | No | Not measured |
| Jitter cleaner | LMK03200 | X-ray | 149 | No | 1% |
| LVDS-LVTTL clock buffer | CDCLVC1310 | X-ray | 12300 | No | 15% |
| LVDS block buffer | CDCLVD1212 | $^{60}$Co $\gamma$ | 11300 | No | 50% |

Table 4 the SEE test result summary

| Device | Radiation type | Non-SEFI SEU | | SEFI | |
|---|---|---|---|---|---|
| | | $\sigma$ (cm$^2$) | Est. err rate (1/week) | $\sigma$ (cm$^2$) | Est. err rate (1/week) |
| AFBR-79EIDZ RX | Proton | <1.3×10$^{-11}$ | <0.02 | <1.3×10$^{-11}$ | <0.02 |
| AFBR-79EIDZ TX+RX | Neutron | <2.3×10$^{-11}$ | <0.03 | <2.3×10$^{-11}$ | <0.03 |
| LMK03200 | Neutron | <2.5×10$^{-11}$ | <0.04 | 6.1×10$^{-10}$ | 0.88 |
| CDCLVD1212 | Neutron | <2.3×10$^{-11}$ | <0.03 | <2.3×10$^{-11}$ | <0.03 |
| CDCLVC1310 | Neutron | <8.0×10$^{-12}$ | <0.01 | <8.0×10$^{-12}$ | <0.01 |
| XC7K325T RX | Neutron | <2.2×10$^{-10}$ | <0.31 | <2.0×10$^{-10}$ | <0.31 |

## 4. Conclusion

A COTS-based clock distribution system has been developed for the ATLAS LAr LTDB Demonstrator. The performance of the clock distribution system has been evaluated. The components used in the clock distribution system have been qualified to meet radiation tolerance requirements of the Demonstrator.

## Acknowledgments

This work is supported by US-ATLAS R&D program for the upgrade of the LHC, the US Department of Energy Grant DE-FG02-04ER1299 and Hubei Provincial Natural Science Foundation of China (Grant Number 2014CFC1093). The authors would like to express the deepest appreciation to Ms. Tanya Herrera, Dr. Steve Wender, and Dr. Ron Nelson, Dr. Helio Takai, Dr. Mike Wirthlin, Mr. Long Huang Dr. Ethan Casio for beneficial discussions and kind help during the irradiation tests.